\documentstyle[twocolumn,psfig]{jpsj}
\def\runtitle{Low-temperature dependence of the London penetration depth}
\def\runauthor{Takao {\sc Morinari} and Manfred {\sc Sigrist}}

\title{The influence of chiral surface states on the London penetration
depth in Sr$_2$RuO$_4$}  

\author{Takao {\sc Morinari} and Manfred {\sc Sigrist}}
\inst{Yukawa Institute for Theoretical Physics, Kyoto University,
Kyoto 606-8502, Japan}

\recdate
{\today}

\abst
{The London penetration depth for the unconventional superconductor
Sr$_2$RuO$_4$ is analyzed assuming an order parameter which breaks time
reversal symmetry and parity simultaneously. Such a superconducting
state possesses chiral quasiparticle states with subgap energies at
the surface. We show that these subgap states can give a significant
contribution to the low-temperature behavior of the London penetration
depth yielding a $ T^2 $ power-law even though bulk quasiparticle
spectrum is gapped. The presence of several electron bands gives rise
to interband transition among the subgap surface states and influences
the properties of the surface impedance. Furthermore, the surface
states lead also to a non-linear Meissner effect. }

\kword
{
Sr$_2$RuO$_4$, $p$-wave superconductor, time-reversal
breaking state, chiral surface state}

\begin{document}
\sloppy
\maketitle

Several years of intense experimental research have established
the unconventional nature of superconductivity in 
Sr$_2$RuO$_4$ \cite{MAENO1,NATRICE}. This compound has a layered
perovskite structure representing a basically two-dimensional metal with
three almost cylindrical Fermi surfaces.  
The symmetry of the superconducting state is very likely 
odd in parity, which implies the spin-triplet configuration analogous
to superfluid $^3$He \cite{RICE,ISHIDA1}. Muon
spin rotation experiments provide evidence for broken 
time reversal symmetry \cite{LUKE}, a fact that strongly suggests that 
the gap function has the basic form,
\begin{equation}
{\bf d} ({\bf k}) = \Delta_0 \hat{{\bf z}} \frac{k_x \pm i k_y}{k_F}
\label{chiralpwave}
\end{equation}
which is a chiral $p$-wave state, here written in the vector
representation, assuming cylindrical symmetry. The Cooper pairs possess 
an internal orbital angular momentum which is oriented along the
$z$-axis. A consequence of this topological property of the
superconducting phase is the presence of chiral surface states at the
surface \cite{VOLOVIK,MATSU,YASU}. 
While the chiral $p$-wave state has a basically gapful quasiparticle
spectrum, the surface states correspond to subgap quasiparticle
excitations with a continuous spectrum down to zero energy
\cite{MATSU,YASU}. These quasiparticle states are Andreev bound states
and extend only over a coherence length towards the bulk. 
In this letter we consider the contribution of these states to the
temperature dependence of the London penetration depth. The London
penetration depth $ \lambda_{\parallel} $ for currents within the
plane and the in-plane coherence length $ \xi_{\parallel} $ are very
similar giving a Ginzburg-Landau parameter $ \kappa =
\lambda_{\parallel}/  \xi_{\parallel} \approx 2.6 $. Therefore, the
presence of the surface 
states can lead to a visible reduction of the screening effect and
could even 
dominate the low-temperature behavior $ \lambda_{\parallel} $, in
particular, if the bulk quasiparticle spectrum is gapped. 
We show here that power-law temperature dependence can result from
the surface states, which is usually taken as an evidence for
nodes in the bulk quasiparticle gap.

The discussion of the London penetration depth requires a careful
analysis of the current-current response to an external field which can
be written in general as
\begin{equation}
j_\mu ({\bf r},t) = - \frac{c}{4 \pi}\sum_{\nu} \int dt' \int d^3r'
K_{\mu \nu}({\bf r},t; {\bf r}',t')  A_\nu ({\bf r}',t')
\end{equation}
where only the transverse component of $ A_\nu ({\bf r}',t') $ enters.
The kernel $ K_{\mu \nu}({\bf r},t; {\bf r}',t') $ is obtained from the
current-current correlation function. In our
case this response consists of two contributions: the bulk part due to
the continuum of quasiparticle states above the gap and the part due
to the surface states. We consider from now on the specific case of a
surface with normal vector along the $x$-axis and an external field
parallel to the $z$-axis. Consequently we have to deal with the
transverse vector potential and screening current along the
$y$-axis. The relevant terms are then,
\begin{equation}
j_y ({\bf r},t) = - \frac{1}{c}\int d^4r' \left[ 
\Pi_{yy} (r; r') 
- \frac{c^2 \delta^{(4)}(r- r')}{4 \pi \lambda_0^2} \right] 
A_y (r')
\label{eq_j}
\end{equation}
where the integral runs of the four coordinates $ r'= ({\bf r}', t') $
and $\lambda_0 $ corresponds to the bare ``London penetration depth''
of the bulk regime which can be considered as basically
temperature-independent for very low temperatures. For this bulk part
we take the local approximation, while for the first part connected
with the surface states nonlocality is important, as we will see
below. 

The surface states can be easily described within the Bogolyubov-de
Gennes formalism if we neglect self-consistency of the gap $ \Delta_0 $
which we choose to be constant everywhere inside the superconductor.
For the sake of simplicity we assume also that the surface provides
specular reflection of quasiparticles and the gap has no anisotropy on
the Fermi surface. The electron band in our model has cylindrical
symmetry and is represented by the parabolic form,
$ \varepsilon_{\bf k} = \{(k_x^2 + k_y^2) - k_{\rm F}^2 \}  / 
2m  $ neglecting any dispersion along the $z$-axis. 
Using these simplifications the wave function of the subgap states
localized at the surface is given by 
\begin{equation}
\left( \begin{array}{c}
u_{\bf k} \left( {\bf r} \right) \\
v_{\bf k} \left( {\bf r} \right)
\end{array} \right) \simeq  \sqrt{\frac{2}{\xi_0L_y d}} ~ {\rm
e}^{ik_y y - \frac{x}{\xi_0}}~ \sin (k_x x)
\left( \begin{array}{c}
1 \\
-i \end{array} \right).
\label{uv}
\end{equation}
Since only states very close to Fermi surface are important for the
low-temperature properties the wave vector
can be represented essentially as $ (k_x , k_y) =k_F (\cos \theta,\sin
\theta)$  for $ k_F \xi_0 \gg 1 $. Further, 
$ L_y $ is the extension of the system along the $ y $-direction with
periodic boundary conditions and $
d$ is the interlayer spacing (the wave function is renormalized per
layer). The energy of the surface states 
is given by $E_{k_y} = \eta \Delta_0 {k_y}/k_F$ with $
\eta = \pm 1 $ denoting the sign of the chirality of $ {\bf d} ({\bf
k}) = \hat{{\bf z}} (k_x \pm i k_y)/k_F $ \cite{HONER}. 
Defining the current operators as $ j_{\mu} = (\hbar e/2mi) 
(\hat{\Psi}^{\dag} \partial_{\mu} \hat{\Psi} - h.c.) $, where
\begin{equation}
\hat{\Psi} ({\bf r})=\left( \begin{array}{c}
\psi_{\uparrow} ({\bf r}) \\ 
\psi_{\downarrow}^{\dag} ({\bf r}) 
\end{array} \right) = \sum_{\bf k} \left[
\begin{array}{cc} u_{\bf k} ({\bf r}) & -v_{\bf k}^* ({\bf r}) \\
v_{\bf k} ({\bf r}) & u_{\bf k}^* ({\bf r}) \end{array} \right]
\left( \begin{array}{c} \gamma_{{\bf k} \uparrow} \\ \gamma_{{\bf
k}\downarrow}^{\dag} \end{array} \right),
\label{N_f}
\end{equation}
is the Nambu field operator with $\gamma^{({\dag})}_{{\bf k}\sigma}$
the Bogolyubov quasi-particle operator, we can express the
current-current correlation function $\Pi_{yy} (r;r')$ as
\begin{eqnarray}
\Pi_{yy} (r;r')
&=& -\frac{\hbar^2 e^2}{4m^2 } \lim_{r_1,r_2 \rightarrow r,~r_1',r_2'
\rightarrow r'} \nonumber \\
& & \hspace*{-15mm}\times (\partial_{y_1} - \partial_{y_2})
(\partial_{y_2'} - \partial_{y_1'})
{\rm Tr} \left[ {\bf G}(r_1; r_1') {\bf G}(r_2'; r_2) \right]
\label{j_j}
\end{eqnarray}
where ${\bf G} (r;r')$ is the Nambu-Gor'kov Green's
function in real space and $ \partial_y $ denotes the derivative with
respect to the spatial $ y$-coordinate.
The Green's function can be expressed as
\begin{equation}
{\bf G} ({\bf r},{\bf r}^{\prime};i\omega_n)
\simeq \frac{\phi (x) \phi (x^{\prime})}{L_y d} \sum_{0<k<k_F}
\sum_{s=\pm} \frac{{\hat{\sigma}}_0 - s {\hat{\sigma}}_2}{i\omega_n -
s E_k} {\rm e}^{isk(y-y^{\prime})}
\label{green_fn}
\end{equation}
with $\phi (x) = \sqrt{2/\xi_0}~ \exp^{-x/\xi_0} \sin k_F x$,
${\hat{\sigma}}_0$ the unit matrix and
${\hat{\sigma}}_2$ the second Pauli matrix,
and $\omega_n$ being the fermionic Matsubara frequency.
Using Eqs. (\ref{j_j}) and (\ref{green_fn}), we calculate
the current-current correlation function. The translational invariance 
along the $ y $-direction allows us to transform the $ y$-coordinate
into momentum space, 
\begin{eqnarray} 
\lefteqn{\Pi_{yy} (x,x^{\prime};q,i\Omega_n)} \nonumber \\
& & \quad \simeq -\frac{32\hbar^2 e^2}{m^2L_y d } g(x) g(x^{\prime})  
\sum_{0< k <k_F} k^2 
\frac{f(E_{k+q})-f(E_k)}{i\Omega_n - E_{k+q}+E_k}  \nonumber \\ 
& & \quad \simeq  - \frac{8\pi \hbar^2 k_F^3}{3d m^2 \Delta_0} 
\left( \frac{k_B T}{\Delta_0} \right)^2 
\frac{g(x) g(x^{\prime})}{1-i\Omega_n k_F /q \Delta_0},
\label{pi}
\end{eqnarray}
in the limit $ T \ll T_c $, 
where $g(x)\simeq \exp(-\frac{2x}{\xi_0}) \sin^2 (k_F x)/ \xi_0 $ is
the square of the amplitude of the surface state wave function and $ i
\Omega_n $ is the bosonic Matsubara frequency.
In deriving Eq. (\ref{pi}), we restrict ourselves to the leading
contribution for $q \ll k_F$ and $k_B T \ll \Delta_0$. 
The nonlocal nature of response enters via the product form $ g(x)
g(x') $ which accounts for the fact that each of the 
quasiparticle state is localized at the surface. The field at the
particular point $ (x',y', z') $  couples to the surface state in the
same layer with a weight $ g(x') $ and yields consequently a response
at any other point $ (x,y,z') $ with weight $ g(x) $. This is a
feature of the effectively one-dimensional character of the surface
states within each layer. A local approach in this place would 
underestimate the role of the surface states in the low-temperature
response.  

Combining Eqs. (\ref{eq_j}) and (\ref{pi}), where we further use the
analytic continuation $i\Omega_n \rightarrow \hbar \omega +
i\delta$, with the Maxwell equation 
$\nabla^2 A_y ({\bf r},\omega) = -\frac{4\pi}{c} j_y ({\bf r},\omega)$
we obtain an integro-differential equation for $A_y ({\bf
r},t)$. The boundary condition is given by  
$\partial_x A_y ({\bf r},\omega)|_{x=0} =
B_z(q,\omega)$, where $ B_z(q,\omega)$ is the external magnetic field
at the surface parallel to the $ z$-axis. We solve this equation using
an approximation $g(x) \simeq \exp(-2x/\xi_0)/2 \xi_0 $ in the
integrand (we ignore the fast oscillations), which is certainly valid
for $k_F \xi_0 \gg 1$. This allows us to calculate the surface
impedance, $Z(q,\omega) = 4\pi c E_y(x=0,q,\omega)/B(q,\omega) = -
4\pi i \omega A_y (x=0,q,\omega)/B(q,\omega)$. We then obtain the
penetration depth using the relation $4\pi \omega \lambda (q,\omega) =
{\rm Im} Z(q, \omega)$. Taking a static limit $\hbar \omega
k_F/q\Delta_0 \rightarrow 0$ and $q \rightarrow 0$, we find in the
regime of $k_B T \ll \Delta_0$, $\Delta \lambda (T) = \lambda(T) -
\lambda(0) $ has $T^2$-behavior:
\begin{equation}
\Delta \lambda (T)/\lambda_0 \simeq \frac{4\pi^2}{3}
\frac{\kappa}{(2\kappa+1)^2} \left( k_B T/\Delta_0 \right)^2.
\label{lambda}
\end{equation}
Setting $\kappa \simeq 2.6$, which is a typical value of
Sr$_2$RuO$_4$, we obtain $\Delta \lambda(T)/\lambda_0 \simeq 0.14
\times \left( T/T_c \right)^2$, if we assume the weak-coupling
relation $ \Delta_0 = 1.76 k_B T_c $. We ignored the temperature
dependence of $ \lambda_0 $ as it is exponential in the
low-temperature regime in our model.

\begin{figure}
\begin{center}
\psfig{file=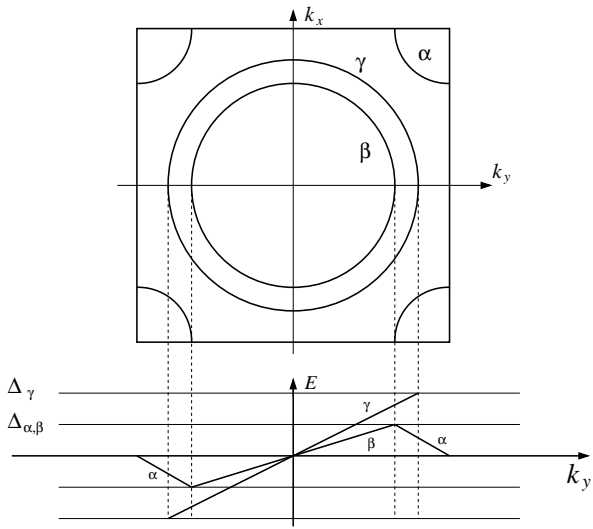,height=7.0cm}
\end{center}
\caption{Schematic spectrum of the chiral surface states in the
multi-band case of Sr$_2$RuO$_4$. The Brillouin zone contains 
three Fermi surfaces. The electron-like $ \beta $
and $ \gamma $-Fermi surfaces yield a chiral surface state spectrum
centered around $ k_y=0 $, while the spectrum of the surface state due
to the hole-like $ \alpha $-Fermi surface located at the Brillouin zone
boundary with opposite sign of chirality. The gap magnitudes are in
general different. We assigned, however, the same magnitude for $
\alpha $ and $ \beta $-Fermi surface for simplicity.}   
\end{figure}

It is important to notice that this contribution is independent of
the sign of the chirality and the charge of the carriers, i.e., whether 
the superconducting state is $ {\bf d} \propto \hat{{\bf z}} (k_x + i
k_y)  $ or $ \hat{{\bf z}}(k_x
- i k_y) $) and the Fermi surface is electron- or hole-like. Therefore, 
the 
formation of domains of the two superconducting states would not
lead to a significant change of the result. 
Furthermore, for the 
case of several superconducting bands the contributions of the surface 
states of each band add up to enlarges the prefactor of the $ T^2
$-law. The coherence length 
as the extension of the surface states towards the interior is
different for each band, since the Fermi velocities and the gap
magnitudes are different. The coherence length,  experimentally
determined via the measurement of $ H_{c2}$, giving $ \kappa \simeq
2.6 $ is the shortest among all. Therefore, the enhancement of the
surface state contribution can be sizable in the multiband case. We
consider here the case of three bands, as schematically shown in
Fig.1, where each band yields its own surface state described by a
Green's function ${\bf G}^{(j)} (r;r')$, (the superscript $j$ labels
the $j^{th}$band). We assume that the reflection of quasiparticles on
the surface does not lead to transitions among the different
bands. Each band is characterized by a Fermi vector $ k^{(j)}_{F} $,
the effective band mass $ m_j $ and the superconducting gap $\Delta_j
$. We now use Eq.(\ref{j_j}) to calculate the contribution of each
band to the current-current correlation function and then analyze the
resulting equation for the transverse vector potential as in the
single band case. This leads to the low-temperature behavior of the
London penetration depth, 
\begin{equation} 
\frac{\Delta \lambda (T)}{\lambda_0} \simeq \frac{4\pi^2}{3} 
\sum^3_{j=1} \frac{\kappa^{(j)}}{(2 \kappa^{(j)}+1)^2}
\left(\frac{\lambda_0 \Delta_0}{\lambda^{(j)} \Delta_j}\right)^2 
\left( \frac{k_B T}{\Delta_0} \right)^2
\label{lambda2}
\end{equation}
where 
\begin{equation}
\frac{1}{\lambda_0^2} = \sum_j \frac{1}{\lambda^{(j)2}} = \sum_j
\frac{2 e^2 k^{(j)2}_F}{m_j c^2 d} 
\label{eq_eta}
\end{equation}
and $ \kappa^{(j)} = \lambda_0 / \xi^{(j)}_0 $ with $ \xi^{(j)}_0 =
\hbar^2 k^{(j)}_F / m_j \Delta_j $. We choose $ \Delta_0 $ to
reproduce again the weak coupling relation, $ \Delta_0 = 1.76 k_B T_c
$. The contribution of all three bands can easily give a prefactor to
the $ (T/T_c)^2 $-law of order one, consistent with recent
measurements for fields along the $ z $-axis.

In the discussion of the multi-band situation we 
neglected the interband effects. Oscillatory fields, for example
appearing in microwave experiments, can yield interband
transitions. The matrix elements for the transition depends on various 
details of the orbital and band structure. We do not go into these
complex details here, but assume that the interband transition can be
described by an ordinary current operator, $ j^{(i,j)}_\mu = (\hbar
e/2m'i) (\hat{\Psi}^{(i)\dag} \partial_\mu \hat{\Psi}^{(j)} - h.c.)$
where $ m' $ is a phenomenological parameter accounting for the matrix
element. $\hat{\Psi}^{(i)}$ is the Nambu field operator of the $i$-th
band as in Eq. (\ref{N_f}). In the small-momentum $(q)$ and
small-frequency $ (\omega) $ limit only surface states are important
which share the same zero-energy momentum. As shown in Fig.1 this is
the case for the $ \beta $- and $ \gamma $-band. The $ \alpha $-band
is unimportant because for interband transitions a large momentum
transfer ($ q \sim \pi $) is  necessary. Analogous to Eq.(\ref{pi}) we
can derive the correlation function, 
\begin{equation} \begin{array}{l} \displaystyle 
\Pi^{(\beta, \gamma)}_{yy}(x_1,x_2,q; i \Omega_n)  
\simeq \frac{16 \hbar^2 e^2}{m'^2d L_y } \tilde{g}(x_1) \tilde{g}(x_2)
\sum_{k} k^2 \\ \\ \displaystyle 
\times \left [ \frac{f(E^{\beta}_{k+q}) - f(E^{\gamma}_k)}{i\Omega_n - 
E^{\beta}_{k+q} + E^{\gamma}_k}+  \frac{f(E^{\gamma}_{k+q}) -
f(E^{\beta}_k)}{i\Omega_n - E^{\gamma}_{k+q} + E^{\beta}_k} \right]
\end{array}
\label{ib-pi}
\end{equation}
where $ \tilde{g}(x) \approx \exp(-2 x / \tilde{\xi})/2
\sqrt{\xi^{(\beta)} \xi^{(\gamma)}} $ with $ \tilde{\xi}^{-1} =
\xi^{(\beta)-1} + \xi^{(\gamma) -1} $. The surface state spectra in the
two bands is  approximated by $ E^{(i)}_k = v_i k $, with $ v_i =
\Delta_i/k^{(i)}_F $. We now take the analytic continuation $ i
\Omega_n \to \hbar \omega + i \delta $ and set $ q=0 $ in
Eq.(\ref{ib-pi}). For the limit of very small $ \hbar \omega $ ($ \ll
k_B T $) the surface impedance gets the following contribution from
the interband transitions, 
\begin{equation} \begin{array}{l} \displaystyle
Im Z_{ib}(\omega) = \frac{4 \pi^3 \tilde{\xi} \xi^{(\gamma)}}{3
\lambda^{(\gamma)2}} \frac{\tilde{\kappa}^3}{(2 \tilde{\kappa}+1)^2} 
\left(\frac{\Delta_0 m_{\gamma}}{\Delta_\gamma m'}\right)^2
\frac{\omega v_{\gamma}}{\tilde{v}} \\ \\
 \displaystyle
\qquad \times \left[ \left(\frac{k_B T}{\Delta_0} \right)^2
\left(\frac{v_{\gamma}^2}{v_{\beta}^2}-1 \right)
+ \frac{2}{\pi} \left(\frac{\hbar \omega}{\Delta_0} \right)^2 {\rm ln} 
\frac{v_{\gamma}}{v_{\beta}} \right]\end{array}
\end{equation}
for the imaginary part which yields in the zero frequency-limit also
a $ T^2 $-contribution to the London penetration depth ($
\tilde{\kappa} = \lambda / \tilde{\xi} $ and $ \tilde{v} = v_{\gamma}
- v_{\beta} > 0 $). For the real part we obtain,
 \begin{equation} \begin{array}{l} \displaystyle
Re Z_{ib}(\omega) = \frac{2 \pi^3 \tilde{\xi} \xi^{(\gamma)}}{
\lambda^{(\gamma)2}} \frac{\tilde{\kappa}^3}{(2 \tilde{\kappa}+1)^2} 
\left(\frac{\Delta_0 m_{\gamma}}{\Delta_\gamma m'}\right)^2
\\ \\  \displaystyle
\qquad \times \left(\frac{v_{\gamma}}{\tilde{v}} \right)^3
\omega \left(\frac{\hbar
\omega}{\Delta_0} \right)^2 \frac{\hbar \omega}{k_B T} 
\end{array}
\end{equation}
for given $ T $ and small $ \hbar \omega $ ($ \ll k_B T \ll \Delta_0 $).
The imaginary part, the inductive resistance, shows a $ \omega
$-linear plus $ \omega^3 $-behavior, while the real part, the surface
resistance follows an $ \omega^4 $-law.  
In the opposite limit where $ k_B T \ll \hbar \omega \ll \Delta_0 $ the
surface impedance due to interband transitions has to vanish. The
reason is that for $ q =0 $ the initial and final states are either
both empty or occupied in the zero-temperature limit. A simple
analysis shows the following behavior,
\begin{equation} \begin{array}{l} \displaystyle
Re Z_{ib}(\omega) \propto \frac{\omega^4}{T} ~{\rm e}^{ -
\frac{v_{\gamma}+ v_{\beta}}{2 \tilde{v}} \frac{\hbar \omega}{k_B T}},
~Im Z_{ib}(\omega) \propto \frac{T^4}{\omega^2}
\end{array} \end{equation}
The surface resistance and the inductive resistance vanish
exponentially and with a power-law, respectively, in the
zero-temperature limit. 

We now consider the possibility of a so-called nonlinear Meissner
effect.The application of a magnetic field 
introduces a Doppler shift which changes the quasiparticle energy,
\begin{equation}
E'_{\bf k} = E^{(sg)}_{\bf k} + \frac{e v_{Fy}}{c} A_y 
\end{equation}
if we again consider the case of $ {\bf n} = (1,0,0) $. We can expand
the current-current correlation function for small $ A_y $ and analyze
the contribution to the London penetration depth in the same way as
done above. Restricting to the single band model we obtain,
\begin{equation} \begin{array}{l} \displaystyle
\frac{\Delta \lambda(T)}{\lambda_0} \simeq \frac{4 \pi^2}{3}
\frac{\kappa}{(2 \kappa +1)^2} \left( \frac{k_B T}{\Delta_0} \right)^2 
\left( 1 - \eta \frac{3 \kappa}{2} \frac{H}{H_{c2}}
\right)
\label{nonlin-pi}
\end{array} \end{equation}
which leads to a non-linear correction in
the external field. This effect is a
consequence of the angular momentum of the Cooper pairs coupling to
the field along the $ z$-axis. The sign of this correction
depends on the chirality (direction of angular momentum) and the
character of the Fermi surface, i.e.,  
a different sign appears for the $ \alpha $-band than for the $ \beta
$- and $ \gamma $-band in Fig.1. Therefore,
the presence of electron-like and hole-like Fermi surfaces as well as
the formation of domains of the two chiral states lead to compensations
which diminish the effect.

In recent experiments by Bonalde et al., the temperature
dependence of the 
London penetration depth was determined 
using a self-inductive technique\cite{BONALD}. 
It was found that the low-temperature behavior is 
indeed governed by a $ T^2 $-behavior. This is not compatible
with a simple interpretation in terms of line nodes in the gap as
originally proposed based on the $ T $-power laws in specific heat and 
NQR \cite{NISHI}, since this would lead to a linear $ T $-dependence
\cite{MINEEV}. The more sophisticated approach based on a nonlocal
response theory by Kostin and Leggett (KL) (for the reason that $
\kappa $ is small), however, would yield a $ T^2 $-behavior
\cite{KOSTIN}. On the other hand, in this letter we propose an
alternative mechanism for a $ T^2$-behavior based on the contributions
of the surface states. In both theories it is expected that this
power-law behavior is absent for $ \lambda_{\perp} $, for screening
currents flowing along the $ z$-axis. The $z$-axis current is not
proportional to the surface state energy as required to obtain the $
T^2 $-behavior in our theory. Furthermore, $ \kappa_{\perp} $ is about
20 times larger than the in-plane $ \kappa $ so that contribution of
the surface states as well as the nonlocal effect by KL are rather
small. However, the measurements for fields in the plane, probing the
$ z $-axis current show a similar $ T^2 $-behavior. This is in
apparent conflict with both interpretations. Since in this case,
however, not only $ \lambda_{\perp} $ but also the contribution from
in-plane currents from the surfaces normal to the $z$-axis are involved,
the final answer will be given only when these geometrical aspects
have been thoroughly investigated \cite{VANHARL}. 

In our model the gap size is isotropic on the Fermi surface.
Anisotropy in turn modifies the surface state spectrum to $ E_{k_y} =
v k_y + v' k_y^3 + ... $ without destroying the particle-hole symmetry
($ v v' < 0 $ in general). This yields an additional $ T^4
$-contribution which may not be so small. Together with other
correction in this order this leads to 
\begin{equation}
\frac{\Delta \lambda(T)}{\lambda_0} = a (T/T_c)^2 + b (T/T_c)^4 + ...
\end{equation}
with $ a \sim b >0 $.
In an intermediate range the second term generates a
T-dependence which over some temperature range appears to be close to
a $ T^3$-behavior and only at rather low temperatures the $ T^2 $-law
would dominate. For one sample Bonalde et al. could indeed fit their
data reasonably well with a $ T^3 $-curve \cite{VANHARL}. While this
sample happened to be dirty it is not clear from the experiment what
is the intrinsic origin for the apparently different
power-law. Therefore, the additional $ T^4 $-contribution, which due
to the surface orientation and disorder is larger than usual, may be
one possible explanation. 

A further aspect noteworthy here is that our mechanism is active at the
surface only, while the KL scheme also applies also in the bulk of the
superconductor. Therefore, the London penetration 
depth governing the magnetic interaction between vortices should have
different temperature dependence in the two scenarios. Measurements of 
in the mixed phase by $ \mu $SR suggest that London penetration 
depth saturates faster than $ T^2  $ at low temperature \cite{MSR}. 
In contrast to the surface-sensitive experiment mentioned above
\cite{BONALD}, $\mu$SR is indeed a bulk probe. 
Unfortunately, it has less accuracy in determining the
temperature dependence of the London penetration depth so that we
cannot draw a strong conclusion to date. Nevertheless,
the present experimental result is consistent with the interpretation
based on surface states. 

We would like to thank A. Furusaki, M. Matsumoto, T.M. Rice,
C. Honerkamp, J. Goryo,
Y. Maeno, D. Van Harlingen and I. Bonalde for many
stimulating discussions. 
This work was supported by a Grant-in-Aid of the Japanese Ministry of
Education, Science, Culture and Sports.

\end{document}